\documentclass[showkeys,showpacs,superscriptaddress]{revtex4}
\usepackage{amsmath}
\usepackage{amssymb}
\usepackage{graphicx}
\usepackage{color}
\usepackage{rotating}
\newcommand{\be}{\begin{equation}}
\newcommand{\ee}{\end{equation}}
\newcommand{\bea}{\begin{eqnarray}}
\newcommand{\eea}{\end{eqnarray}}

\begin{document}

\title{Nonperturbative quantization \`{a} la Heisenberg \\and thermodynamics of monopole configurations}

\author{
Vladimir Dzhunushaliev
}
\email{v.dzhunushaliev@gmail.com}

\affiliation{Department of Theoretical and Nuclear Physics, Kazakh National University,\\
Almaty 050040, Kazakhstan}
\affiliation{
	IETP, Al-Farabi Kazakh National University, Almaty 050040, Kazakhstan
}
\affiliation{
	Institute of Physicotechnical Problems and Material Science of the NAS of the Kyrgyz Republic, 265 a, Chui Street, Bishkek 720071,  Kyrgyz Republic
}
\affiliation{
	Institute of Systems Science,
	Durban University of Technology, P. O. Box 1334, Durban 4000, South Africa
}

\author{Vladimir Folomeev}
\email{vfolomeev@mail.ru}
\affiliation{
	IETP,  Al-Farabi Kazakh National University, Almaty 050040, Kazakhstan
}
\affiliation{
	Institute of Physicotechnical Problems and Material Science of the NAS of the Kyrgyz Republic, 265 a, Chui Street, Bishkek 720071,  Kyrgyz Republic
}

\author{Hernando Quevedo}
\email{quevedo@nucleares.unam.mx}

\affiliation{Instituto de Ciencias Nucleares, Universidad Nacional Aut\'{o}noma de M\'{e}xico,\\
AP 70543, Ciudad de M\'{e}xico 04510, M\'{e}xico}

\affiliation{	Dipartimento di Fisica and ICRA, Universit\`a di Roma ``La Sapienza'',  Piazzale Aldo Moro 5, I-00185 Roma, Italy
}

\affiliation{Department of Theoretical and Nuclear Physics, Kazakh National University,\\
Almaty 050040, Kazakhstan}

\begin{abstract}
For field theories in which no small parameter is available, we propose a definition of nonperturbative quantum states in terms of the complete set of Green functions,
based upon the utilization of  Heisenberg's quantization procedure. We apply this method to obtain the energy spectra of a quantum monopole and of a flux tube.
The partition function and thermodynamic quantities for one quantum monopole  are evaluated numerically.
 A dilute gas of noninteracting  quantum monopoles is considered, for which  the partition function and thermodynamic quantities are evaluated as well.
 All the obtained statistical and thermodynamic functions contain quantum corrections associated with the internal structure of the quantum monopole.
\end{abstract}

\pacs{12.38.Lg; 12.90.+b; 11.15.Tk}

\keywords{
Nonperturbative quantization, quantum monopole, monopole gas, thermodynamics
}
\date{\today}

\maketitle

\section{Introduction}

In theoretical physics, it is now widely believed that the physical world has an intrinsic quantum nature. It should therefore be possible to develop an exact quantum description of all physical systems.  However, this is not the case. Indeed, a quantum-mechanical description implies solving the Schr\"odinger equation for a wave function. Simple mechanical systems can be treated in this way, and idealized solutions can be found analytically. Nevertheless, more realistic physical systems imply the direct numerical solution of Schr\"odinger's equation, using a finite-element approximation for the operators involved in the corresponding differential equation. In general, this method is very ineffective due to the large number of resulting variables. It is therefore necessary to replace the physical system by an idealized model which admits an exact solution. Then, one tries to determine a perturbation series for the perturbation operator which is determined by the difference between the Hamiltonian of the physical system and the Hamiltonian of the idealized model. This approach can be applied only if there exist small parameters with respect to which the perturbation operator can be expanded. Most physical systems, however, do not permit the existence of such a small parameter that could be used in the framework of perturbation theory. Moreover, it can happen that the small parameter exists, but the perturbation series has a vanishing convergence radius. In all these cases, a nonperturbative approach is necessary in order to describe the quantum properties of the physical system.

An essential component of perturbative quantization is the Hilbert space of quantum states which is constructed as follows. The perturbative approach allows us to introduce the creation and annihilation operators  $		\hat a^\dagger \text{ and } \hat a$, respectively. Then, the vacuum state $ \left| vac \right\rangle $ is introduced by means of the condition
$
\hat a \left| vac \right\rangle  = 0$. Analogously, a quantum state with $n$ particles is defined as
$\left| n\right\rangle = (\hat a^\dagger)^n \left| vac\right\rangle$. The Hilbert space is understood as the sum of all $n-$particle states
$\left| n\right\rangle$. We see that the concept or particle is essential to introduce the Hilbert space. If there is no small parameter, the physical system cannot be represented in terms of particles and, consequently, no Hilbert space can be defined in the standard manner. This is an important fact that must be considered in the construction of any alternative nonperturbative approach to quantization.
		
The problem of handling physical systems with no small parameter has been treated in many different ways. For instance, the direct numerical integration of the Schr\"odinger equation or the corresponding field equations, in which an auxiliary parameter (the step of integration) is introduced, has been used intensively. In this connection, lattice models are commonly used in quantum field theory; in this case, the lattice constant, which is the auxiliary nonphysical parameter used to calculate functional integrals, determines the accuracy of the lattice approach. Numerical methods encounter usually technical problems due to the exponential growth of calculations. Moreover, physical systems with a large number of degrees of freedom lead to instabilities of the numerical approaches.

Nonnumerical approaches include the Hartree-Fock method, the procedures for approximating Hamiltonians, the density functional theory, etc.  Most of these nonnumerical methods can solve the main problem of estimating the physical properties of the ground state.
However, successive states cannot be calculated in general with the required accuracy and
the determination of the spectrum of the entire system is a difficult task in most systems \cite{strocchi13}.
Some of these difficulties are overcome by the operator method which reduces all the differential calculations to algebraic calculus with the matrix elements of the operators.
The eigenfunctions and eigenvalues of the Hamiltonian are easily calculated in the zeroth approximation, whereas the successive approximations lead to convergent sequences so that physical quantities can be estimated with any desirable accuracy \cite{filu15}.
In the context of supersymmetric quantum field theory, the orbifold equivalence and other methods  have been applied to understand the quark confinement in gauge theories. These investigations have also provided some progress towards the formulation of a nonperturbative definition of the path integral in quantum field theory \cite{dunn16}.

Although there have been several attempts to construct a consistent approach to nonperturbative quantization and many interesting technical results have been obtained, no definite approach has been formulated so far. In view of this situation, we consider that it is convenient to explore alternative methods which could shed some light on the complexity of the problem. In this work, we follow this strategy and propose a definition of nonperturbative quantum states which can be used in a consistent manner to find certain characteristics of quantum systems. This new definition is based upon the use of  Green functions in Heisenberg's quantization procedure.  The main idea is simple. Since the Green functions are solutions of the corresponding operator field equations, they should contain all the quantum information about the physical system. To exemplify our method we will study in this work the thermodynamics of a quantum monopole and of a dilute monopole gas.

The main difference between thermodynamics of  perturbatively quantized fields and  thermodynamics of nonperturbatively quantized fields is that in the first case
a physical system can be enclosed in an arbitrary volume filled with particles -- quanta (for example, photons). In the second case the fields can create
 self-supporting objects with fields exponentially decaying at infinity. These objects are protons, neutrons, nuclei, glueballs, etc.
In the first case, calculation techniques for determining the partition function are well known. In the second case, the energy of such a physical system is
practically concentrated in a restricted region of space without any walls. Linear sizes and the volume of this region are determined by some field parameters.

In QCD lattice calculations, it was shown with great certainty that the QCD vacuum has a complex structure:
it contains magnetic monopoles \cite{Nambu:1974zg,Hooft,Mandelstam:1974pi,Ripka:2003vv}.
A great deal of works has been devoted to studies of the monopoles in different aspects, including
the problem of confinement in QCD, the problem of proton decay, in cosmology and astrophysics, etc.
(see, for example, an extensive literature on the subject in the book~\cite{Shnir:2005xx}). In the context of the present paper, we just make a brief mention of some of these works
devoted to a monopole condensate and a dilute gas of monopoles.
In Refs.~\cite{DiGiacomo:1999yas,DiGiacomo:1999fb}, the dual superconductivity of the vacuum in the $SU(3)$ gauge theory is investigated by constructing a disorder parameter which signals monopole condensation.
In those studies, there was used the  Abelian projection method, within which, as some believe, it is possible to get
monopoles that are more relevant than others for confinement.
Numerical studies of Abelian monopoles in  lattice gauge theory are presented in Refs.~\cite{Chernodub:1997dr,Chernodub:1997ay}.
In turn, a study of $SU(2)$ dilute monopole gas was initiated by
 Polyakov's paper~\cite{Polyakov:1976fu}, where, within the framework of
the semiclassical quantization of $SU(2)$ monopole solutions,  the effect of the Debye screening by a dilute gas of monopoles was demonstrated.
Developing this idea,  the dilute gas of monopoles was considered in different  aspects (including the thermodynamic ones)
in Refs.~\cite{Nesti:1996rm,Martemyanov:1997ks,Agasian:1997wv,Chernodub:2000mi,Davis:2001mg,Chernodub:2004qp,Das:2009fb}.
These studies include a consideration of magnetic monopoles in a fully nonperturbative way in lattice Monte Carlo simulations~\cite{Davis:2001mg}.
In turn, the authors of Ref.~\cite{Chernodub:2004qp}  come to the conclusion that even though
the Abelian monopole gas in the 3-dimensional $SU(2)$ gluodynamics is not dilute,
the dilute monopole gas approximation is adequately consistent with the measurements of the monopole density and the Debye screening mass.

Consistent with all this, the purpose of the present paper is to demonstrate the fact that there is a possibility to study thermodynamics of a dilute gas of monopoles within the nonperturbative quantization \`{a} la Heisenberg. To do this, we first examine thermodynamics of a single quantum monopole obtained by one of us earlier in Ref.~\cite{Dzhunushaliev:2017rin}, after which we compute thermodynamic functions of the dilute gas of monopoles.

The physical system under investigation is a quantum condensate  filled with quantum monopoles. The quantum monopoles are supported by those degrees of  freedom of the  $SU(3)$
gauge field that belong to the  $SU(2)$ subgroup of the $SU(3)$ gauge field, i.e. $SU(2)\subset SU(3)$.
The quantum condensate is described by the coset degrees of  freedom $SU(3)/SU(2)$.

There is a crucial difference between these degrees of  freedom: the $SU(2)$ fields have a nonzero quantum average with quantum fluctuations around this average. The dispersion of these quantum fluctuations is approximately described as some constant determined from solving a nonlinear eigenvalue problem. To find quantum averages of the $SU(2)$ field, we use the Yang-Mills equation corresponding to the $SU(2)$ subgroup. In turn, the quantum condensate is supported by the coset space gauge fields, which have a zero quantum average and nonzero dispersion, and whose expectation value over all color and spacetime indices is described by the scalar  field $\phi$.

To describe the physical system under consideration, we employ the two-equation approximation of Refs.~\cite{Dzhunushaliev:2017rin,Dzhunushaliev:2016svj},
within which the Yang-Mills and scalar-field equations are solved as a nonlinear eigenvalue problem. Having such a solution, we calculate the energy spectrum of a quantum monopole
that enables us to evaluate a partition function for a single quantum monopole and also to get all corresponding thermodynamic functions.

In computing the partition function of the monopole gas we take into account the internal structure of monopoles
that results in the appearance of an extra term in the expression for the energy of the physical system under investigation. To simplify the difficult problem of determining the energy
of the system of quantum monopoles, we assume that the motion of monopoles (and, hence, their kinetic energy) can be represented as the motion of point-like particles.

This work is organized as follows. In Sec. \ref{sec:qs}, we use Heisenberg's quantization to propose a definition of nonperturbative quantum states in terms of Green functions. Some properties of this definition are also discussed. We also present the  explicit equations in   the case of an $SU(3)$ gauge field.
In Sec.~\ref{pfm}, within the framework of the two-equation approximation, we construct the energy spectra of
  a single monopole  and of a flux tube which can connect monopole and antimonopole in the vacuum of QCD. Using these solutions, we evaluate numerically the partition function
  of a single monopole and calculate the corresponding thermodynamic quantities.
  In Sec.~\ref{pfgm} we consider a nonrelativistic dilute gas of noninteracting monopoles, for which we find the partition function and derive the resulting thermodynamics.
  In Sec.~\ref{connQCDpsi} we demonstrate the relationship between one of free parameters of our system and the typical energy scale of QCD. Finally, in
Sec.~\ref{concl} we summarize the obtained results.


\section{Nonperturbative quantum states}
\label{sec:qs}

The absence of a small parameter in a physical system prevents it from being quantized perturbatively and, consequently, the Hilbert space from being defined. This means that alternative approaches should be considered in order to characterize nonperturbative quantum states which, accordingly, cannot be considered as vectors of a Hilbert space. As an alternative approach, we will consider in the following analysis Heisenberg's quantization method \cite{hei66}. To be more specific, consider a Lagrangian density functional ${\cal L}(\Phi^A)$ for a set  of fields $\Phi^A(x)$. The variational principle leads to the field equations
\be
	D {\cal L} \equiv \frac{\partial}{\partial x^\mu} \left(\frac{\partial {\cal L}}{\partial \Phi^A_{,\mu}}\right)
	- \frac{\partial {\cal L} }{\partial \Phi^A} = 0 \ .
\ee

According to Heisenberg's procedure, the first step towards the nonperturbative quantization of a classical system   consists in replacing the classical fields $\Phi^A$ by the corresponding operators $\hat{\Phi}^A$. Then, the quantum counterpart of the above field equations is
\be
	D\hat{\cal L} = 0\ , \qquad \hat{\cal L} =   \hat{\cal L}(\hat{\Phi}^A).
\label{opeq}
\ee
In general, there is no method to solve this kind of operator equation. To avoid this difficulty, one can alternatively consider the average of the operator equation over all possible products of the field operator $\Phi^A$, i. e., one  considers the entire set of  Green functions which are determined by means of the set of equations
\bea
	\left\langle D\hat{\cal L} \right\rangle & = & 0 \ ,
\nonumber \\
	\left\langle \hat \Phi^A(x_1) D\hat{\cal L} \right\rangle & = &0 \ ,
\nonumber \\
	\left\langle \hat \Phi^{A_1}(x_1) \hat \Phi^{A_2}(x_2) D\hat{\cal L} \right\rangle &  =&  0 \ ,
\nonumber \\
	\cdots & = & 0 \ ,  \\
	\left\langle \hat \Phi^{A_1}(x_1) \hat \Phi^{A_2}(x_2)\cdots \hat\Phi^{A_n}(x_n) D\hat{\cal L} \right\rangle &  =&  0 \ ,
\nonumber \\
	\cdots & = & 0 \ .
\nonumber
\eea
This represents an infinite set of differential equations that must be solved for any particular physical system; if this can be done, we end up with an infinite set of Green functions which should contain all the physical information about the field operators and the quantum states of the system. \emph{It then follows that we can identify the set of nonperturbative quantum states with the set of Green functions.} This is the main observation that allows us to formally define nonperturbative quantum states in terms of Green functions.

Of course, solving an infinite set of equations is not possible, unless we come up with a method to recursively solve the equations or to truncate the equation series. In general, this is not an easy task. Nevertheless, the situation is not hopeless since examples can be found in which this procedure leads to consistent results. Consider, for instance, the case of a linear field theory. In this case, it is possible to calculate the propagator; then, all the Green functions can be represented as polylinear combinations of the propagator. In other words,  Heisenberg's nonperturbative approach is simply equivalent to the canonical quantization of linear systems in terms of propagators. We thus see that our definition of nonperturbative quantum states in term of Green functions is trivially realized in the case of linear field theories.

An important consequence of defining quantum states as above is that the classical limit,
which is usually a difficult task in most nonperturbative approaches,  can easily be handled. Indeed, if all Green functions can be represented as products of the field functions, then the system is classic. Consequently, to obtain the classical limit, it is necessary to represent the complete set of Green functions in terms of standard field functions. This can be considered as a nontrivial advantage of our
definition of nonperturbative quantum states.

In the case of gauge theories, nonperturbative quantum states defined in terms of Green functions can also be handled explicitly.
For the sake of concreteness, let us consider an $SU(3)$ gauge field. Then, the operator equation (\ref{opeq}) leads to the Yang-Mills equations
\begin{equation}
	D_\nu \widehat {F}^{A \mu\nu} = 0
\label{2-10}
\end{equation}
where
\be
	\hat F^B_{\mu \nu} =
	\partial_\mu \hat A^B_\nu - \partial_\nu \hat A^B_\mu +
	g f^{BCD} \hat A^C_\mu \hat A^D_\nu
\ee
is the field strength operator; $\hat A^B_\mu$ is the gauge potential operator; $B, C, D = 1, \ldots , 8$ are the $SU(3)$ color indices;
$g$ is the coupling constant; and $f^{BCD}$ are the structure constants for the $SU(3)$ gauge group.

The Yang-Mills operator equations (\ref{2-10}) are equivalent to an infinite set of equations for the Green functions, i. e.,
\begin{eqnarray}
	\left\langle
		D_\nu \widehat {F}^{A \mu\nu} (x)
	\right\rangle &=& 0 ,
\nonumber \\
	\left\langle
		\hat A^{B_1}_{\alpha_1} (x_1)
		D_\nu \widehat {F}^{A \mu\nu} (x)
	\right\rangle &=& 0 ,
\nonumber\\
	\left\langle
		\hat A^{B_1}_{\alpha_1} (x_1) \hat A^{B_2}_{\alpha_2} (x_2)
		D_\nu \widehat {F}^{A \mu\nu} (x)
	\right\rangle &=& 0 ,
\label{ymopeq}\\
	\ldots &=& 0	,
\nonumber\\
	\left\langle
		\hat A^{B_1}_{\alpha_1} (x_1) \ldots \hat A^{B_n}_{\alpha_n} (x_n)
		D_\nu \widehat {F}^{A \mu\nu} (x)
	\right\rangle &=& 0
\nonumber\\
	\ldots &=& 0	.
\nonumber
\end{eqnarray}
This system possesses different particular solutions each of which determines a particular quantum state. The physical properties of each state must be investigated separately; however, one can expect that some states will correspond to  standard $SU(3)$ configurations.

A construction similar to that described above was used in Refs.~\cite{Dzhunushaliev:2015mva,Dzhunushaliev:2013nea} to demonstrate that modified
gravity theories arise as a consequence of applying the nonperturbative  quantization procedure to general relativity.

\section{Energy spectrum and partition function of a quantum monopole in contact with a thermostat}
\label{pfm}

In order to demonstrate how one can calculate thermodynamic quantities for nonperturbatively quantized fields, we employ the simplest object supported by such fields -- a quantum monopole~\cite{Dzhunushaliev:2017rin}.

\subsection{Energy spectrum of a quantum monopole}

In general, it is not possible to solve the entire set of equations (\ref{ymopeq}); however, one can consider the average quantum values as given by the classical value of the potential plus a fluctuation term that can be used to truncate the set of equations.  Indeed, this procedure has been performed in \cite{Dzhunushaliev:2016svj}, leading to the result that Eqs.(\ref{ymopeq}) reduce to a set of two equations, namely,
\begin{eqnarray}
	\tilde D_\nu F^{a \mu \nu} - \left[
		\left( m^2 \right)^{ab \mu \nu} -
		\left( \mu^2 \right)^{ab \mu \nu}
	\right] A^b_\nu &=& 0 ,
\label{1-1-10}\\
	\Box \phi - \left( m^2_\phi \right)^{ab \mu \nu} A^a_\nu A^b_\mu \phi -
	\lambda \phi \left( M^2 - \phi^2  \right) &=& 0,
\label{1-1-20}
\end{eqnarray}
where $\tilde D_\mu$ is the gauge derivative of the subgroup $SU(2)$;
$\left( m^2 \right)^{ab \mu\nu}$,
$\left( \mu^2 \right)^{ab \mu \nu}$, and $\left( m^2_\phi \right)^{ab \mu \nu}$ are quantum corrections coming from the dispersions of the operators
$\widehat{\delta A}^{a \mu}$ and $\hat A^{m \mu}$:
\begin{eqnarray}
	\hat A^{a \mu} &=& \left\langle \hat A^{a \mu} \right\rangle +
	i \widehat{\delta A}^{a \mu} ,
\label{1-1-30}\\
	\left\langle \hat A^{m \mu} \right\rangle &=& 0 .
\label{1-1-40}
\end{eqnarray}
The quantum averaging  $\left\langle \ldots \right\rangle $ entering the formulae
\eqref{1-1-30} and \eqref{1-1-40} is understood to mean the averaging over the nonperturbative quantum state defined in Sec.~\ref{sec:qs}.
Recall that such a quantum state is determined by the entire set of  Green functions. In this case, the nonperturbative quantum state is approximately determined by the
2-point Green functions \eqref{1-1-50}, \eqref{1-1-60} and the 4-point Green function  \eqref{1-1-85} defined below. 

The quantum corrections $\left( m^2 \right)^{ab \mu\nu}$,
$\left( \mu^2 \right)^{ab \mu \nu}$, and $\left( m^2_\phi \right)^{ab \mu \nu}$ are determined in terms of the 2-point Green functions
\begin{eqnarray}
	G^{mn \mu \nu}(y,x) &=& \left\langle
		\hat A^{m \mu}(y) \hat A^{n \nu}(x)
	\right\rangle ,
\label{1-1-50}\\
	G^{ab \mu \nu}(y,x) &=& \left\langle
		\widehat{\delta A}^{a \mu}(y) \widehat{\delta A}^{b \nu}(x)
	\right\rangle .
\label{1-1-60}
\end{eqnarray}
We approximate these functions as follows
\begin{eqnarray}
	G^{mn \mu \nu}(y,x) &\approx&
	- \Delta^{mn} \mathcal A^\mu \mathcal A^\nu \phi(y) \phi(x) ,
\label{1-1-70}\\
	G^{ab \mu \nu}(y,x) &\approx&
	\Delta^{ab}	\mathcal B^\mu \mathcal B^\nu ,
\label{1-1-80}
\end{eqnarray}
where $\Delta^{ab} (a,b=2,5,7), \Delta^{mn} (m,n = 1,3,4,6,8)$ are constants; 
$\mathcal A_\mu \mathcal A^\nu, \mathcal B_\mu \mathcal B^\nu = \text{const}$.

The 2-point Green function \eqref{1-1-70} requires a more detailed discussion. The point is that the Green function
$G^{mm  \mu \mu}(x,x)$, being the dispersion, is defined in the following manner
\begin{equation}
	G^{mm  \mu \mu}(x,x) = -
	\Delta^{mm} \left( \mathcal A^\mu \right)^2 \phi(x)^2 < 0,
\label{1-1-81}
\end{equation}
and it is a negative quantity. In quantum mechanics, the dispersion is defined by the following known expression
\begin{equation}
	\left\langle \left(
		\hat L - \left\langle L \right\rangle
	\right)^2 \right\rangle =
	\int \psi^* \left(
	\hat L - \left\langle L \right\rangle
	\right)^2 \psi dV > 0
\label{1-1-83}
\end{equation}
and it is a positive quantity, if  the operator $\hat L$ is a Hermitian conjugate operator.
Similarly, in perturbative quantum field theory, the dispersion is also a  positive quantity. But this is not the case in our problem. The point is that, according to our approach, a nonperturbative quantum state, including the dispersion, is defined from solving the infinite set of equations~\eqref{ymopeq}. One may find  that the corresponding solution has a negative dispersion. Such a quantum state may be called a \emph{strange quantum state}. Notice that in our case, if the dispersion \eqref{1-1-81} is positive,  the corresponding solutions obtained below do not exist. Note also that the quantum quantity $\left( m^2_\phi \right)^{ab \mu \nu}$ is associated with the Green function 
$G^{ab \mu \nu}$ and the negativity of the corresponding dispersion is analogous to the appearance  of an imaginary mass for the corresponding quantum in perturbative quantum field theory. We notice that everything stated above applies equally well to a dispersion of the quantity $\hat A^a_\mu$ since we took into account the negativity of the corresponding dispersion by introducing the imaginary unit in the formula \eqref{1-1-30}.

Notice that to derive Eq.~\eqref{1-1-20}, we use the following approximation of the  4-point Green function
\begin{equation}
	G^{mnpq}_{\phantom{mnpq}\mu \nu \rho \sigma}(x, y, z, u) =
	\left\langle
		\hat A^m_\mu(x) \hat A^n_\nu(y) \hat A^p_\rho(z) \hat A^q_\sigma(u)
	\right\rangle,
\label{1-1-85}
\end{equation}
\begin{eqnarray}
    G^{(4)} \approx \frac{\lambda}{4} \left(
        G^{(2)} - M^2  \
    \right)^2 ,
\label{1-1-90}
\end{eqnarray}
where $\lambda, M$ are some constants.
A detailed procedure for obtaining Eqs. \eqref{1-1-10} and \eqref{1-1-20} can be found in Refs.~\cite{Dzhunushaliev:2017rin,Dzhunushaliev:2016svj}.

In order to obtain equations describing the quantum monopole, we employ
the ans\"atz for the $SU(2)$ gauge fields in the standard monopole form in spherical coordinates
\begin{eqnarray}
	A^a_\mu &=& \frac{2}{g} \left[ 1 - f(r) \right]
	\begin{pmatrix}
		0	& 0 & 0 & - \sin^2 \theta \\
		0	& 0 & \cos \varphi & - \sin \theta \cos \theta \sin \varphi \\
		0	& 0 & \sin \varphi & \phantom{-} \sin \theta \cos \theta \cos \varphi \\
	\end{pmatrix} ,
\label{1-1-100} \\
	\phi &=& \frac{\psi(r)}{g},
\end{eqnarray}
where $g$ is the coupling constant. Here $a=2,5,7$ and the spacetime index $\mu = t,r,\theta,\varphi$. Also, we use the following approximation for
$\Delta^{AB} (A,B = 1,2 , \ldots , 8)$:
\begin{equation}
	\Delta^{AB} = \text{diag} \left(
		\Delta_{11}, \delta_{2}, \Delta_{33}, \Delta_{44},
		 \delta_{5}, \Delta_{66}, \delta_{7}, \Delta_{88}
	\right)
\label{1-1-110}
\end{equation}
with $\Delta_{66} = \Delta_{44}$,
$\Delta_{44} + \Delta_{88} = \Delta_{11} + \Delta_{33}$ and the vectors $\mathcal A_\mu, \mathcal B_\mu$
\begin{eqnarray}
	\mathcal A_\mu &=& \left(
		\mathcal A_0, \mathcal A_1, 0, 0
	\right) ,
\label{1-1-120}\\
	\mathcal B_\mu &=& \left(
		\mathcal B_0, 0, 0, 0
	\right)  .
\label{1-1-130}
\end{eqnarray}
As a result, we obtain the following set of equations describing the  quantum monopole:
\begin{eqnarray}
	- f^{\prime \prime} + \frac{f \left( f^2 - 1 \right) }{x^2} -
	m^2 \left( 1 - f \right)\tilde{\phi}^2 &=& - \tilde{\mu}^2 \left( 1 - f \right) ,
\label{1-1-140}\\
	\tilde{\phi}^{\prime \prime} + \frac{2}{x} \tilde{\phi}^\prime &=&
	\tilde{\phi} \left[
		m^2 \frac{\left( 1 - f \right)^2}{x^2} +
		\tilde{\lambda} \left(
			\tilde{\phi}^2 - \tilde{M}^2
		\right)
	\right],
\label{1-1-150}
\end{eqnarray}
where we have introduced the dimensionless variables $x=r/r_0, \tilde\phi=r_0 \psi, \tilde \mu=r_0 \mu, \tilde M= g r_0 M, \tilde \lambda=\lambda/g^2$. Here $r_0$ is a characteristic size of the system. It can be rewritten in terms of some field constant  $\psi_1$ as $r_0 = \psi_1^{-1}$, which will be used below in defining the partition function of the monopole. Note also that, for simplicity, hereafter we set $m_\phi=m$.

\begin{figure}[h]
\begin{minipage}[ht]{.45\linewidth}
		\begin{center}
							\includegraphics[width=1\linewidth]{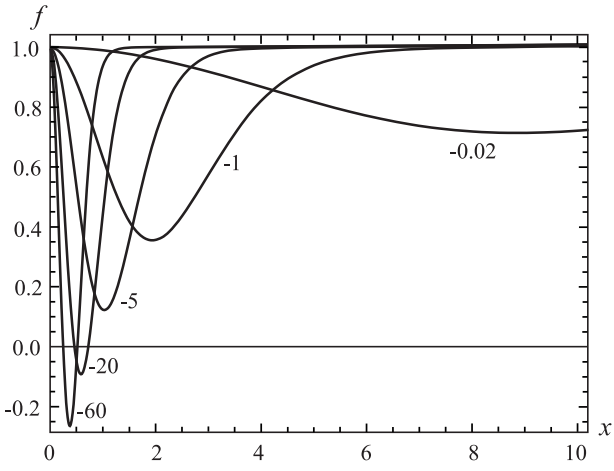}
		\end{center}
			\end{minipage}\hfill
	\begin{minipage}[ht]{.45\linewidth}
		\begin{center}
							\includegraphics[width=1\linewidth]{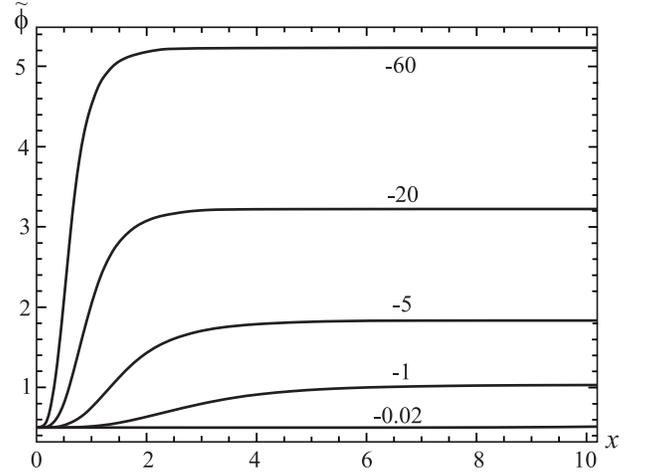}
		\end{center}
			\end{minipage}
\caption{The eigenfunctions  $f(x)$  and $\tilde\phi(x)$ for different values of	$\tilde f_2 = -0.02, -1.0, -5.0, -20.0, -60.0$ and $\tilde\phi_0=0.5, m=2$.
			}
\label{f_phi_plots}	
\end{figure}

The obtained set of equations is solved numerically as a nonlinear eigenvalue problem for the eigenvalues  $\tilde\mu$ and $\tilde M$ and the eigenfunctions
$f(x)$ and $\tilde\phi(x)$. To do this, we choose the following boundary conditions at the center:
\begin{equation}
	f(0) = 1, \quad f^\prime(0) = 0, \quad \tilde\phi(0)\equiv \tilde\phi_0 = \text{const.}, \quad \tilde\phi^\prime(0) = 0 .
\label{1-1-160}
\end{equation}
For numerical solving, it is necessary to have a solution in the neighbourhood of the center, which can be presented as a power series
\begin{eqnarray}
	f(x) &=& 1 + f_2 \frac{r^2}{2} + \ldots =
	1 + \tilde f_2 \frac{x^2}{2} + \ldots ,
\label{1-1-170}\\
	\tilde\phi (x) &=& \tilde\phi_0 +\tilde \phi_2 \frac{x^2}{2} + \ldots ,
\label{1-1-180}
\end{eqnarray}
where the expansion coefficient $f_2$ is arbitrary  (in the dimensionless units, $\tilde f_2 = f_2/\psi_1^2$) and 
$\tilde\phi_2 = \tilde\lambda \tilde\phi_0 (\tilde\phi_0^2 - \tilde{M}^2)/3$.
Using these boundary conditions, we have obtained the families of solutions of Eqs.~\eqref{1-1-140} and \eqref{1-1-150} for different  $\tilde f_2$ shown in Fig.~\ref{f_phi_plots}.

\begin{figure}[h]
	\begin{minipage}[ht]{.45\linewidth}
		\begin{center}
							\includegraphics[width=1\linewidth]{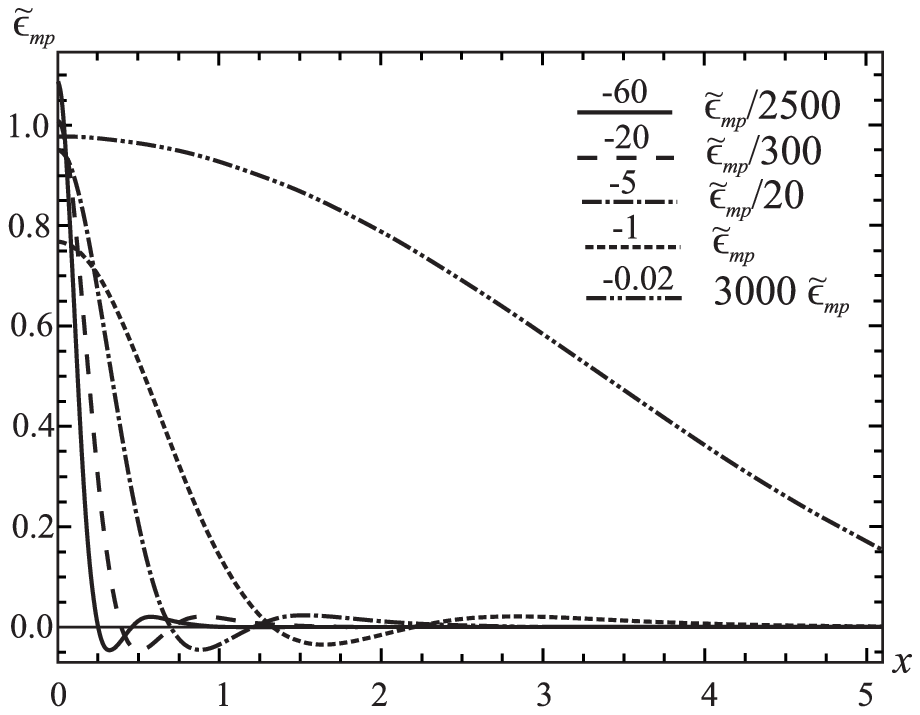}
			\end{center}
\vspace{-.5cm}
		\caption{The dimensionless monopole energy density $\tilde\epsilon_{mp}(x)$ from \eqref{1-1-190}
for different values of $\tilde f_2 = -0.02, -1.0, -5.0, -20.0, -60.0$.
			}
	\label{energ_plots}
	\end{minipage}\hfill
	\begin{minipage}[ht]{.45\linewidth}
		\begin{center}
				\includegraphics[width=1\linewidth]{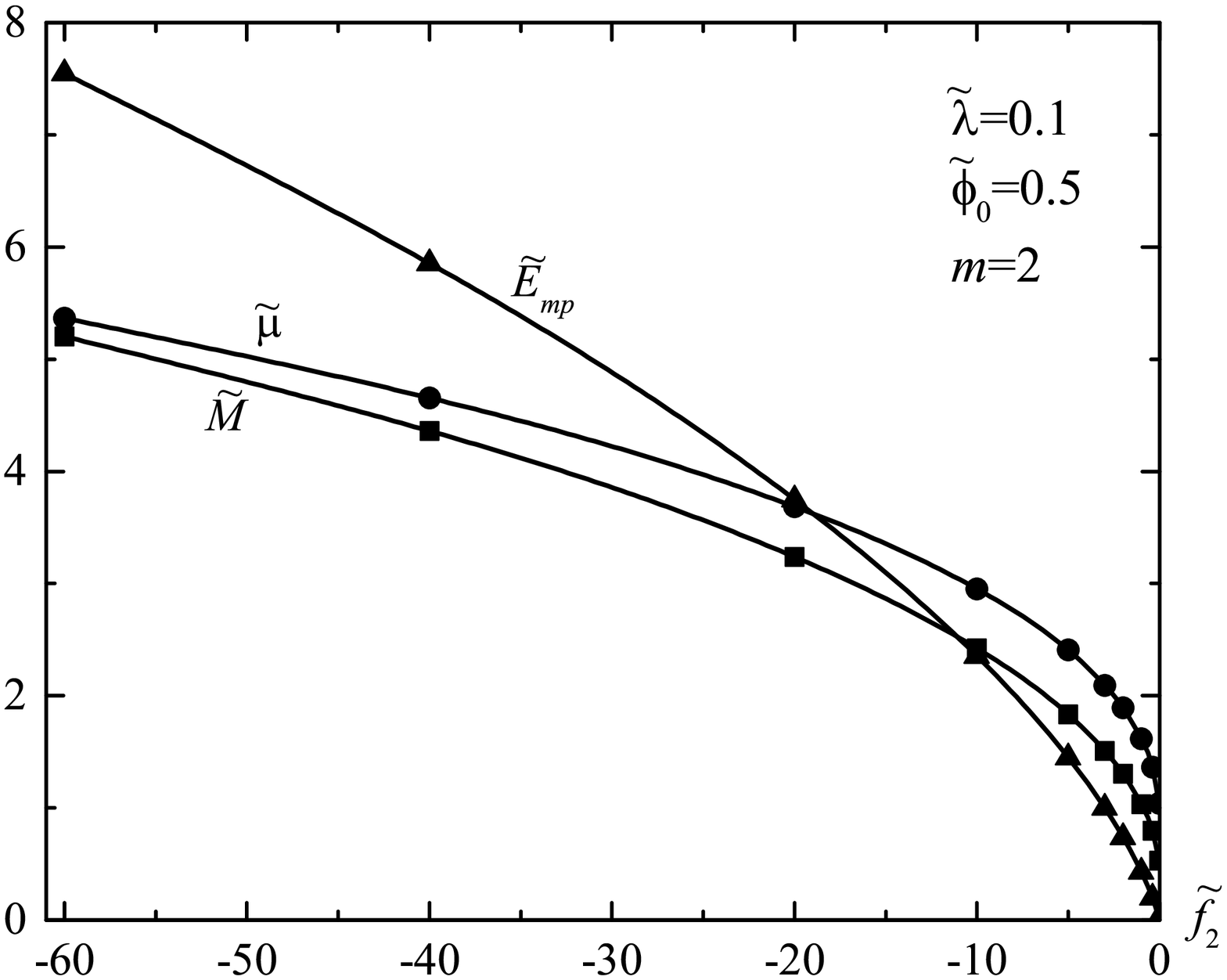}
			\end{center}
\vspace{-.5cm}
		\caption{The dependencies of the monopole total energy $\tilde E_{mp}$ from \eqref{1-1-200}
and of the eigenvalues $\tilde\mu, \tilde M$ on the parameter $\tilde f_2$.
		}
		\label{plots_monop}
	\end{minipage}
\end{figure}

Asymptotic behavior of the eigenfunctions  $f(x), \tilde\phi(x)$ can be found from the analysis of the set of equations \eqref{1-1-140} and \eqref{1-1-150} in the following form
\begin{eqnarray}
	f(x) &\approx& 1 - f_\infty e^{- x \sqrt{m^2 \tilde{M}^2 - \tilde\mu^2}} ,
\label{1-1-184}\\
	\tilde\phi (x) &\approx& \tilde M -
 	\tilde\phi_\infty \frac{e^{- x \sqrt{2 \tilde\lambda \tilde{M}^2}}}{x}.
\label{1-1-188}
\end{eqnarray}
Here $f_\infty, \tilde\phi_\infty$ are constants whose values are determined by the parameter $\tilde f_2$.

The dimensional energy density of the quantum monopole under consideration is given by the expression
\begin{equation}
	L_{mp} = \epsilon_{mp} =\frac{1}{g^2}\left[
\frac{1}{2} \frac{{f'}^2}{ r^2} + \frac{{\psi'}^2}{2} +
	\frac{1}{4 }\frac{\left( f^2 - 1 \right)^2}{r^4} +
	\frac{m^2}{2} \frac{\left( f - 1 \right)^2}{r^2} \psi^2 -
	\frac{1}{2 }\frac{\mu^2}{r^2} \left( f - 1 \right)^2 +
	\frac{\tilde\lambda}{4} \left(
		\psi^2 - g^2 M^2
	\right)^2 + g^2\epsilon_\infty \right],
\label{1-1-190}
\end{equation}
and it is plotted in Fig.~\ref{energ_plots} in the dimensionless form. Using \eqref{1-1-190}, the total energy is calculated as follows:
\begin{equation}
	E_{mp}=4\pi \int\limits_0^\infty \epsilon_{mp} r^2 dr=\frac{4\pi}{g^2 r_0}\int\limits_0^\infty \tilde \epsilon_{mp} x^2 dx=
\frac{4\pi \psi_1}{g^2}\int\limits_0^\infty \tilde \epsilon_{mp} x^2 dx=
\frac{4\pi \hbar c \psi_1}{\tilde{g}^2}\int\limits_0^\infty \tilde \epsilon_{mp} x^2 dx=
\frac{4\pi \hbar c \psi_1}{\tilde{g}^2} \tilde{E}_{mp},
\label{1-1-200}
\end{equation}
where we have introduced the dimensionless coupling constant 
$\tilde{g}^2=\hbar c g^2$; $\tilde \epsilon_{mp}$ and $\tilde E_{mp}$ are the dimensionless energy density and total energy, respectively. Fig.~\ref{plots_monop} shows the dependencies of the total energy and of the eigenvalues  $\tilde\mu, \tilde M$ on the parameter $\tilde f_2$, from which one can see that the limiting values
are
\begin{equation}
	\tilde\mu \xrightarrow{\tilde f_2 \rightarrow 0} 1, \quad
	\tilde M \xrightarrow{\tilde f_2 \rightarrow 0} \tilde\phi_0, \quad
	\tilde E_{mp}\xrightarrow{\tilde f_2 \rightarrow 0} 0 .
\label{1-1-210}
\end{equation}
In Table~\ref{tab1} the eigenvalues $\tilde\mu, \tilde M$ and the energy $\tilde E_{mp}$ are shown as functions of the parameter $\tilde f_2$.

\begin{table}[h]
\scalebox{0.55}{
	\begin{tabular}{|c|c|c|c|c|c|c|c|c|c|c|c|c|c|c|c|c|c|c|c|c|}
	\hline
	\rule[-1ex]{0pt}{2.5ex}
	$\tilde f_2$&-60&-40&-20&-10&-5&-3&-2&-1.6&-1.4&-1.2&-1&-0.9&-0.8&-0.6&-0.4&-0.2&-0.1&
	-0.05&-0.03&-0.02  \\
	\hline
	\rule[-1ex]{0pt}{2.5ex}
	$\tilde M$&5.202&4.36&3.238&2.421&1.833&1.508&1.3029&1.2051&1.152&1.095&1.0334&1&
	0.965&0.8885&0.7982&0.687&0.6123&0.565&0.54228&0.52945  \\
	\hline
	\rule[-1ex]{0pt}{2.5ex}
	$\tilde\mu$&5.364585475&4.65430559&3.68434345&2.95400186&2.406699273&2.09391213&
	1.8908366&1.791790814&1.737779882&1.6793213&1.61522&1.5805559&1.5437766&
	1.4619973&1.364495&1.237047&1.1492464&1.09011125&1.0601184&1.042631\\
	\hline
	\rule[-1ex]{0pt}{2.5ex}
	$\tilde E_{mp}$&7.5457&5.85209&3.75006&2.35408&1.450611&.00007&0.738754&0.622474&0.560275&
	0.496008&0.42951&0.394122&0.35823&0.283069&0.200936&0.11228&0.0607634&
	0.0332751&0.0216438&0.0199801  \\
	\hline
	\end{tabular}
}
\caption{
The eigenvalues $\tilde\mu, \tilde M$ and the energy
$\tilde E_{mp}$ as functions of the parameter $\tilde f_2$ for the monopole.}
\label{tab1}
\end{table}

\subsection{Energy spectrum of a flux tube}

In this section we consider an infinite flux tube filled with a transverse magnetic field. Such a flux tube with a finite length can appear between monopole and antimonopole in the vacuum of QCD.

In order to obtain the corresponding solutions, we use the same two-equation approximation \eqref{1-1-10} and \eqref{1-1-20} and the following  ans\"atz describing the transverse magnetic field creating  an infinite flux tube
\begin{eqnarray}
	A^2_\varphi &=& \frac{\rho}{g} w(\rho),
\label{1-2-10} \\
	\phi &=& \frac{\psi(\rho)}{g}
\label{1-2-20}
\end{eqnarray}
in cylindrical coordinates  $z, \rho, \varphi$. The matrix $\Delta^{AB}$ is chosen in the form \eqref{1-1-110} with arbitrary $\Delta_{11}, \Delta_{33}, \Delta_{44}, \Delta_{66}, \Delta_{88}$ and with the vectors $\mathcal A_\mu, \mathcal B_\mu$
\begin{eqnarray}
	\mathcal A_\mu &=& \left(
		\mathcal A_0, \mathcal A_1, \mathcal A_2, 0
	\right) ,
\label{1-2-30}\\
	\mathcal B_\mu &=& \left(
		\mathcal B_0, \mathcal B_1, 0, 0
	\right) .
\label{1-2-40}
\end{eqnarray}
Using the ans\"atz \eqref{1-2-10}-\eqref{1-2-20} and the auxiliary quantities  \eqref{1-2-30}-\eqref{1-2-40}, we obtain the following equations describing the  infinite flux tube filled with a transverse magnetic field and embedded into a quantum condensate described by the scalar field $\phi$:
\begin{eqnarray}
	- \tilde w'' - \frac{\tilde w'}{x} + \frac{\tilde w}{x^2} + m^2 \tilde\phi^2 \tilde w &=&
	\tilde\mu^2 \tilde w ,
\label{1-2-50}\\
	\tilde\phi^{\prime \prime } + \frac{\tilde\phi'}{x} &=&
	\tilde\phi \left[
		m^2 \tilde w^2 + \tilde\lambda \left(
			\tilde\phi^2 - \tilde M^2
		\right)
	\right].
\label{1-2-60}
\end{eqnarray}
In these equations we have used the following dimensionless variables:
$x=\rho/\rho_0, \tilde w= \rho_0 w, \tilde\phi=\rho_0 \psi, \tilde \mu=\rho_0 \mu$, $\tilde M= g \rho_0 M, \tilde \lambda=\lambda/g^2$.

\begin{figure}[h]
	\begin{minipage}[ht]{.45\linewidth}
		\begin{center}
			\includegraphics[width=1\linewidth]{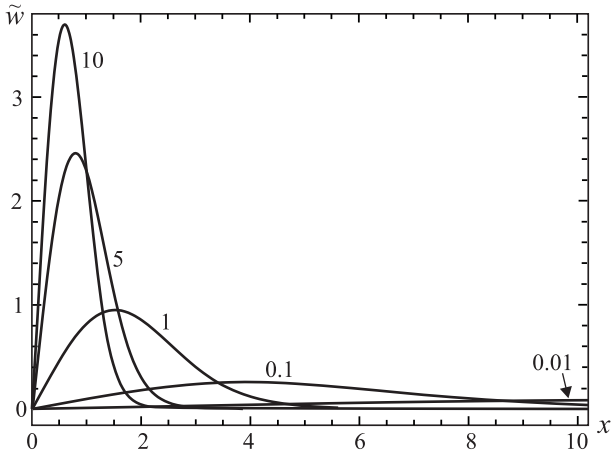}
		\end{center}
	\end{minipage}\hfill
	\begin{minipage}[ht]{.45\linewidth}
		\begin{center}
							\includegraphics[width=1\linewidth]{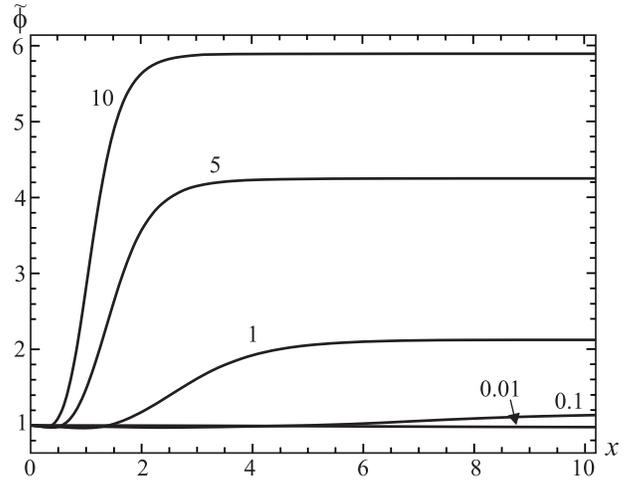}
			\end{center}
		\end{minipage}
\caption{
	The eigenfunctions  $\tilde w(x)$ and $\tilde \phi(x)$  for different values of $\tilde w_1 = 0.01, 0.1, 1.0, 5.0, 10.0$ and $\tilde \phi_0=1, m=1$.
}
\label{w_phi_plots_ft}
\end{figure}

Just as in the previous section, the set of equations \eqref{1-2-50} and \eqref{1-2-60} is solved as a nonlinear eigenvalue problem with the following boundary conditions:
\begin{equation}
	\tilde w(0) = 0, \quad \tilde w'(0) = w_1, \quad \tilde \phi(0)\equiv \tilde \phi_0 = \text{const.}, \quad \tilde \phi'(0) = 0.
\label{1-2-70}
\end{equation}
In order to obtain numerical solutions, we use a Taylor expansion of the functions $\tilde  w(x), \tilde \phi(x)$ at the origin:
\begin{eqnarray}
	\tilde w(x) &=& \tilde w_1 x + \tilde w_3 \frac{x^3}{6} + \ldots ,
\label{1-2-80}\\
	\tilde \phi (x) &=& \tilde \phi_0 +\tilde  \phi_2 \frac{x^2}{2} + \ldots,
\label{1-2-90}
\end{eqnarray}
where $\tilde w_1$ is arbitrary and $\tilde \phi_2 =\tilde  \lambda \tilde \phi_0 (\tilde \phi_0^2 -\tilde  M^2)/2$.
The corresponding numerical solutions are given in Fig.~\ref{w_phi_plots_ft}.

\begin{figure}[h]
	\begin{minipage}[ht]{.45\linewidth}
		\begin{center}
			\includegraphics[width=1 \linewidth]{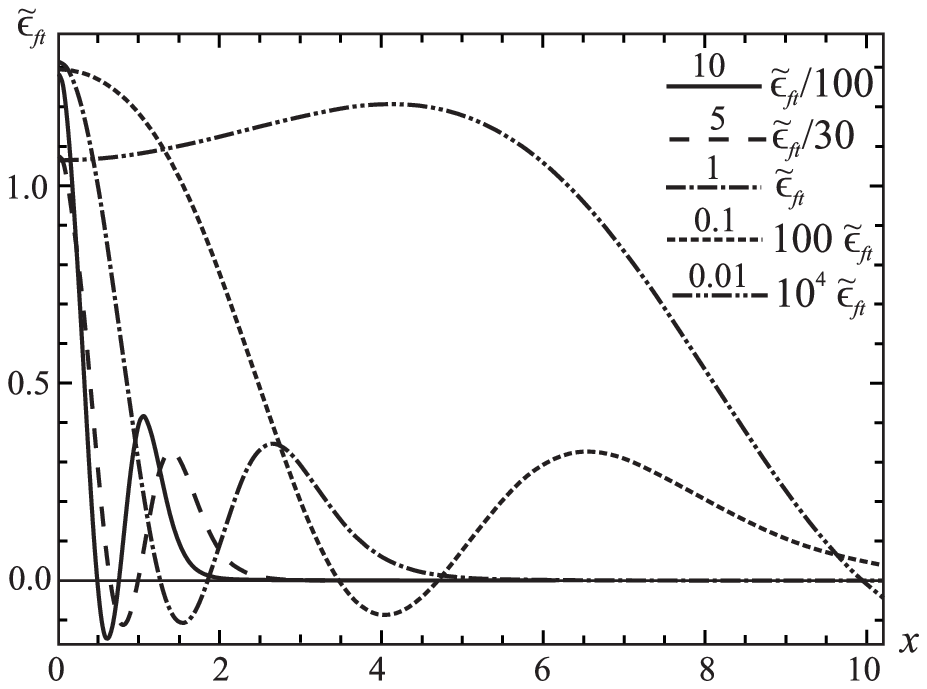}
		\end{center}
\vspace{-.5cm}		
\caption{The dimensionless flux tube energy density $\tilde\epsilon_{ft}(x)$ for different values of $\tilde w_1 = 0.01, 0.1, 1.0, 5.0, 10.0$.
			}
	\label{energ_plots_ft}
	\end{minipage}\hfill
	\begin{minipage}[ht]{.45\linewidth}
		\begin{center}
				\includegraphics[width=1\linewidth]{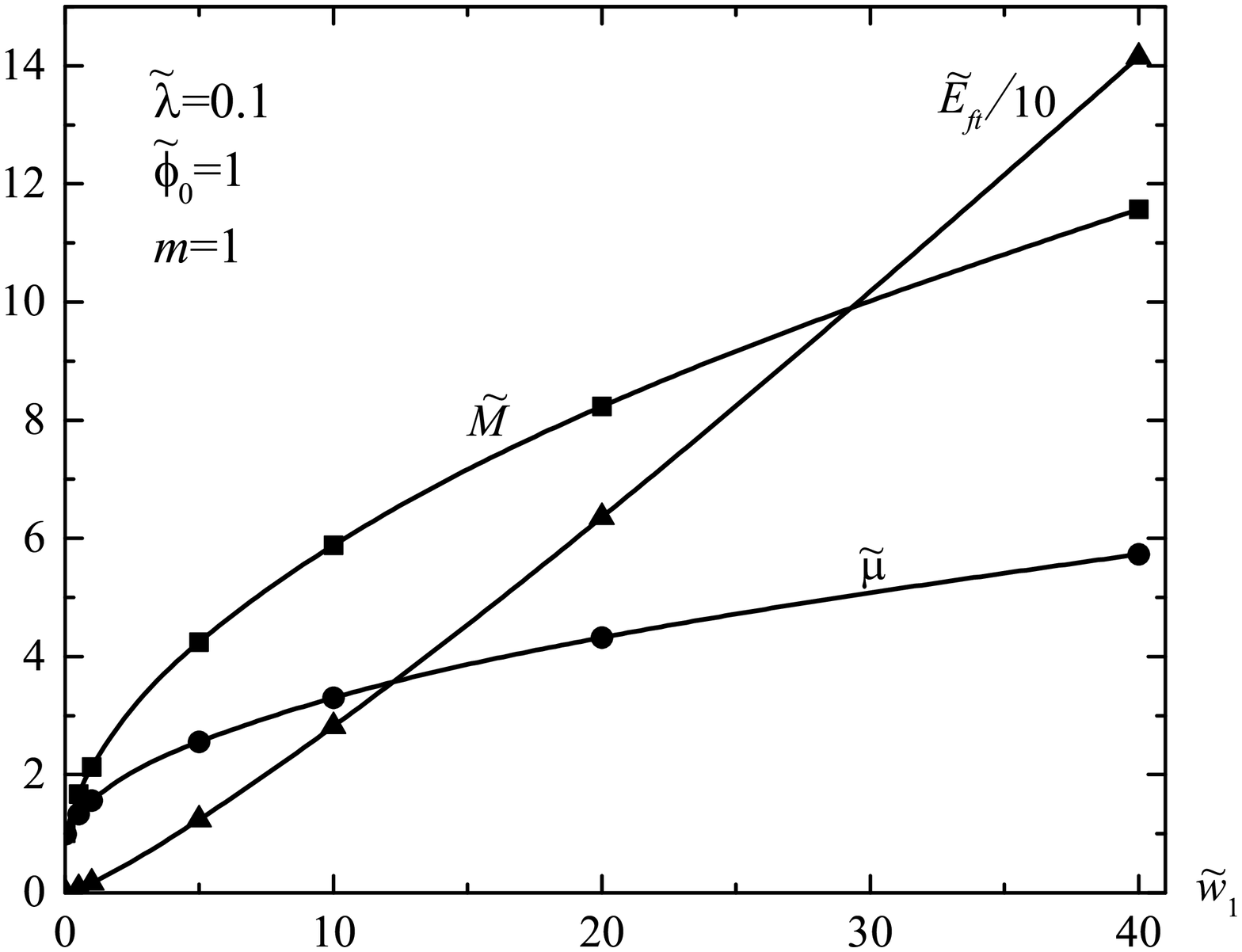}
		\end{center}
\vspace{-.5cm}		
\caption{
	The dependencies of the eigenvalues $\tilde\mu, \tilde M$ and of the linear energy density  $\tilde E_{ft}$ on the parameter	$\tilde w_1$.
		}
		\label{plots_ft}
	\end{minipage}
\end{figure}

Asymptotic behavior of the eigenfunctions  $\tilde w(x), \tilde \phi(x)$ can be found from the analysis of Eqs.~\eqref{1-2-50} and \eqref{1-2-60} in the following form:
\begin{eqnarray}
	\tilde w(x) &\approx& \tilde w_\infty
	\frac{e^{-x \sqrt{m^2 \tilde M^2 - \tilde \mu^2}}}{\sqrt{x}} ,
\label{1-2-100}\\
	\tilde \phi (x) &\approx& \tilde M -
 	\tilde \phi_\infty \frac{e^{- x \sqrt{2 \tilde \lambda \tilde M^2}}}{\sqrt{x}}.
\label{1-2-110}
\end{eqnarray}
Here $\tilde w_\infty, \tilde \phi_\infty$ are constants whose values are determined by the parameter  $\tilde w_1$.

With the above solutions in hand, one can calculate the distribution of
the transverse color magnetic field $H^2_z$, which is defined as follows:
\begin{equation}
	H^2_z = \frac{1}{g}\left(
		w' + \frac{w}{\rho}
	\right).
\label{1-2-120}
\end{equation}
The dimensional flux tube energy density is given by the expression
\begin{equation}
	L_{ft} = \epsilon_{ft} = \frac{1}{g^2}\left[
\frac{{w'}^2}{2} + \frac{{\psi'}^2}{2} +
	\frac{w^2}{2 r^2} + \frac{m^2}{2} w^2 \psi^2 -
	\frac{\mu^2}{2} w^2 +
	\frac{\tilde \lambda}{4} \left(
		\psi^2 - g^2 M^2
	\right)^2 + g^2 \epsilon_\infty\right] ,
\label{1-2-130}
\end{equation}
and it is shown in Fig.~\ref{energ_plots_ft} in the dimensionless form. Using \eqref{1-2-130}, the linear energy density is calculated as follows:
\begin{equation}
	E_{ft} = 2\pi \int\limits_0^\infty \epsilon_{ft} r dr=
\frac{2 \pi }{g^2 \rho_0^2} \int\limits_0^\infty
	 \tilde\epsilon_{ft} x dx =
	\frac{2 \pi \hbar c }{\tilde{g}^2 \rho_0^2} \int\limits_0^\infty
	 \tilde\epsilon_{ft} x dx =
	\frac{2 \pi \hbar c }{\tilde{g}^2 \rho_0^2} \tilde E_{ft},
\label{1-2-140}
\end{equation}
where $\tilde\epsilon_{ft}$ and $\tilde E_{ft}$ are the dimensionless energy density and linear energy density, respectively.
Fig.~\ref{plots_ft} shows the dependencies of the linear energy density and of the eigenvalues  $\tilde\mu, \tilde M$ on the parameter $\tilde w_1$,
from which one can observe the following limiting values:
\begin{equation}
	\tilde\mu \xrightarrow{\tilde w_1 \rightarrow 0} 1, \quad
	\tilde M \xrightarrow{\tilde w_1 \rightarrow 0} \tilde\phi_0, \quad
	\tilde E_{ft} \xrightarrow{\tilde w_1 \rightarrow 0} 0 .
\label{1-2-150}
\end{equation}
Table \ref{tab2} shows the eigenvalues  $\tilde\mu, \tilde M$ and the energy $\tilde E_{ft}$ as functions of the parameter $\tilde w_1$.

\begin{table}[h]
\scalebox{0.8}{
	\begin{tabular}{|c|c|c|c|c|c|c|c|c|c|c|c|c|c|c|}
	\hline
	\rule[-1ex]{0pt}{2.5ex}
	$\tilde w_1$&0.01&0.02&0.04&0.06&0.08&0.1&0.2&0.5&1&2&5&10&20&40  \\
	\hline
	\rule[-1ex]{0pt}{2.5ex}
	$\tilde M$&1.008006233&1.0229746&1.0571195&1.092076&1.126168&1.159194&1.31045&1.67313&2.1319048&2.8212&4.24578&
5.8864985&8.2323&11.5705 \\
	\hline
	\rule[-1ex]{0pt}{2.5ex}
	$\tilde\mu$&0.996365&1.001275&1.017275&1.03485&1.052685&1.070163&
	1.14936&1.3346237&1.5621423&1.8948609&2.559342415&3.2980544&4.31989&
	5.726869\\
	\hline
	\rule[-1ex]{0pt}{2.5ex}
	$\tilde E_{ft}$&0.0099494&0.0198091&0.0399769&0.0618725&0.0848041&0.108471&0.242524&0.737779&1.73539&
	4.07002&12.3594&28.2219&63.5609&141.505  \\
	\hline
	\end{tabular}
	}
\caption{
The eigenvalues $\tilde\mu, \tilde M$ and the linear energy density $\tilde E_{ft}$ as functions of the parameter $\tilde w_1$ for the flux tube.}
\label{tab2}
\end{table}

\subsection{Partition function of the quantum monopole}
\label{2c}

In order to calculate  the partition function of the quantum monopole,
 it is necessary to know on what parameters its energy depends. Then the partition function will be given by an integral over this parameter,
 since in the presence of statistical fluctuations the energy of the system, and hence  the corresponding parameters, are varying.

According to Eq.~\eqref{1-1-200}, the total energy of the quantum monopole consists of finite and infinite parts:
\begin{equation}
	E_{mp} = \frac{4 \pi \hbar c \psi_1}{\tilde{g}^2} \left(
		\int\limits_0^\infty x^2 \tilde \epsilon dx +
		\int\limits_0^\infty x^2 \tilde\epsilon_\infty dx
	\right) = E_1 + E_2.
\label{1-3-10}
\end{equation}
Here the first term,  $E_1$, is finite, and the second one, $E_2$,  is infinite. The total energy $E_{mp}$ depends on the parameters  $f_2$ and $\psi_1$. When statistical fluctuations occur, the  fluctuation of the energy $\delta E_{mp}$ should be finite leading to changes in the parameters $f_2$ and $\psi_1$. But as $\psi_1$ varies, the second integral in \eqref{1-3-10} changes by an infinite amount. It is clear that when the statistical fluctuations are present the thermostat cannot provide an infinite energy, and therefore the parameter $\psi_1$ cannot vary and must remain constant in the presence of the statistical fluctuations.

In general, the partition function is calculated as
\begin{equation}
	Z = \int e^{-\frac{H(\pi_A, \phi^A)}{k T}} \prod d \pi_A d \phi^A,
\label{1-3-20}
\end{equation}
where $H(\pi_A, \phi^A)$ is the Hamiltonian of the system, $\phi^A, \pi_A$ are some generalized coordinates and momenta determining the energy of the system, and $A$ is a collective index for all available indices. Consistent with the above, in our case the energy of the quantum monopole is determined by the parameter   $f_2$. Hence, in the first approximation, the partition function can be calculated in the following way:
\begin{equation}
	Z_{mp}(T) \approx \frac{1}{\psi_1^2} \int e^{-\frac{E_1(f_2)}{k T}} df_2 =
	\int e^{-\frac{\tilde E_1(\tilde f_2)}{\tilde T}} d\tilde{f_2} =
	\tilde Z_{mp} (\tilde T).
\label{1-3-30}
\end{equation}
Here $\tilde E_1$ is the dimensionless energy $E_1$ from \eqref{1-3-10} and 
$\tilde T = k T \tilde{g}^2/(4 \pi \hbar c \psi_1)$ is the dimensionless temperature.

\begin{figure}[t]
	\begin{minipage}[ht]{.45\linewidth}
		\begin{center}
			\includegraphics[width=1\linewidth]{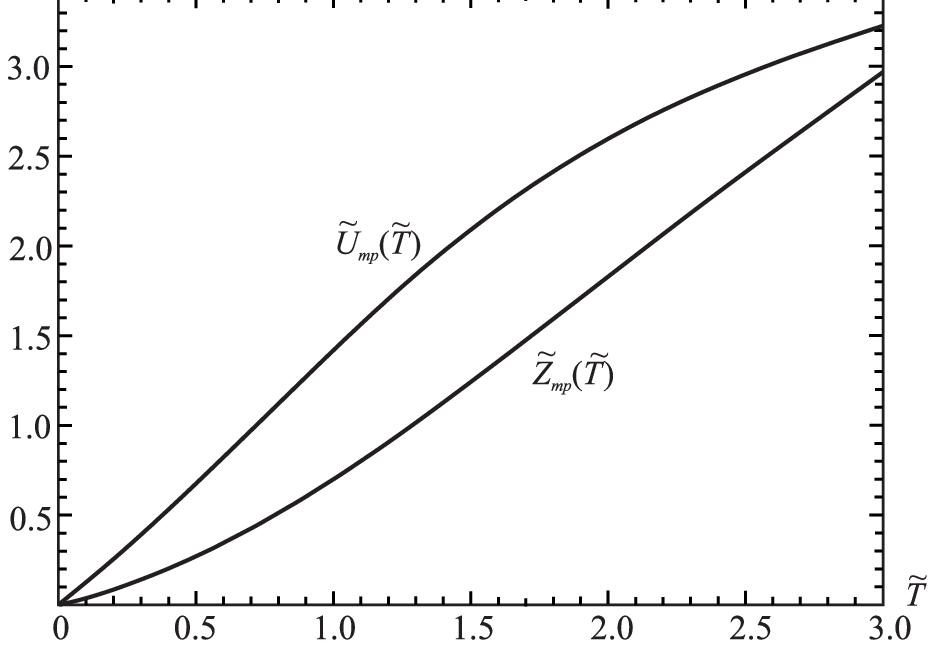}
		\end{center}
		\caption{The dimensionless internal energy $\tilde U_{mp}(\tilde T)$ and the partition function
		$\tilde Z_{mp}(\tilde T)$ of a single monopole.
			}
	\label{inner_energy_partition_fns}
	\end{minipage}\hfill
	\begin{minipage}[ht]{.45\linewidth}
		\begin{center}
			\includegraphics[width=1\linewidth]{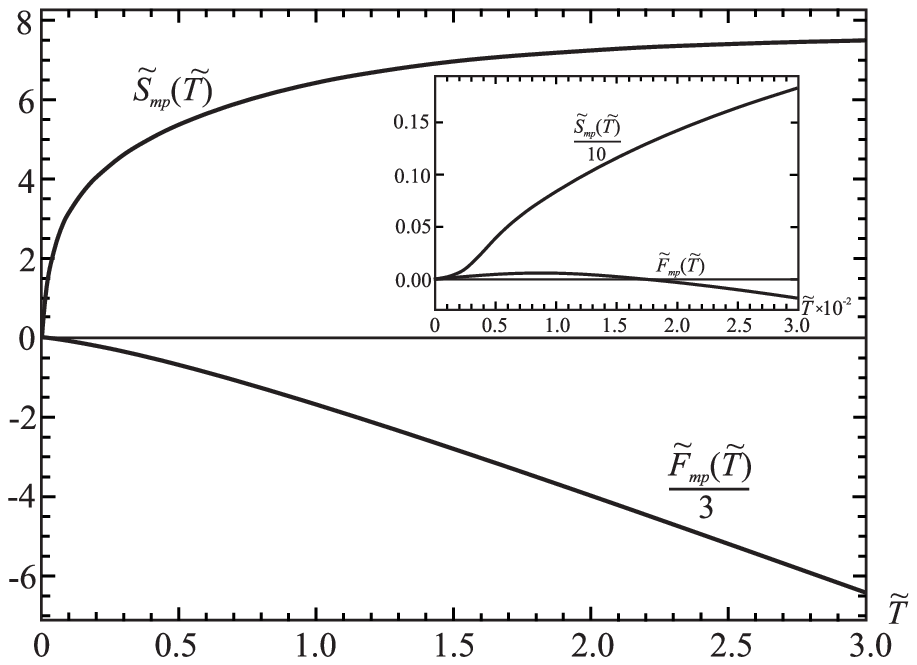}
		\end{center}
\vspace{-0.5cm}		
\caption{
			The dimensionless entropy $\tilde S_{mp}(\tilde T)$ and the Helmholtz free energy $\tilde F_{mp}(\tilde T)$  of a single monopole.
The inset shows the behavior of the functions in the neighbourhood of the center.
		}
		\label{entrp_in_enrg_Helmgolz}
	\end{minipage}
\end{figure}

Having  numerical values of $\tilde E_1$ in hand, we can evaluate the integral \eqref{1-3-30} only for the values of the temperature satisfying the condition
\begin{equation}
	\frac{\tilde E_{1,max}}{\tilde T} \gg 1, \quad
	T \ll \frac{4 \pi \hbar c \psi_1}{k\tilde{g}^2} \tilde E.
\label{1-3-40}
\end{equation}
Here $\tilde E_{1,max}$ is the maximum value of the dimensionless energy \eqref{1-1-200} obtained in  numerical solving Eqs.~\eqref{1-1-140} and \eqref{1-1-150}. For example, for $\psi_1 \approx 10^{15} \text{m}^{-1}$,
which corresponds to the typical length-scale of QCD $r_0\approx 10^{-15}\text{m}$,  we have
\begin{equation}
	T_{max} \approx \frac{4 \pi \hbar c \psi_1}{k\tilde{g}^2} \tilde E \approx
	10^{15} \text{K}.
\label{1-3-50}
\end{equation}
The internal energy is calculated as follows:
\begin{equation}
	U_{mp}(T) = \frac{1}{Z_{mp}} \int E_1(f_2) e^{-\frac{E_1(f_2)}{k T}} d\tilde f_2 =
	\frac{\int E_1(f_2) e^{-\frac{E_1(f_2)}{k T}} d\tilde f_2}
	{\int e^{-\frac{E_1(f_2)}{k T}} d\tilde f_2} =
	\frac{4 \pi \hbar c \psi_1}{\tilde{g}^2}
	\frac{\int \tilde E_1(\tilde f_2)
	e^{-\frac{\tilde E_1(\tilde f_2)}{\tilde  T}} d\tilde f_2}
		{\int e^{-\frac{\tilde E_1(\tilde f_2)}{\tilde  T}} d\tilde f_2} =
	\frac{4 \pi \hbar c \psi_1}{\tilde{g}^2} \tilde U_{mp}(\tilde T) .
\label{1-3-60}
\end{equation}
The dependencies of the dimensionless  partition function  $\tilde Z_{mp}$ from \eqref{1-3-30} and of the dimensionless internal energy $\tilde U_{mp}$ from \eqref{1-3-60} on the  temperature $\tilde T$ are shown in Fig.~\ref{inner_energy_partition_fns}.

The entropy $S_{mp}(T)$ and the Helmholtz free energy $F_{mp}(T) = U_{mp}(T) - T S_{mp}(T)$ are calculated in the following way:
\begin{eqnarray}
	S_{mp}(T) &=& k \left(
		\ln Z_{mp}(T) + \frac{U_{mp}(T)}{k T}
	\right) = k \left(
		\ln \tilde Z_{mp}(\tilde T) + \frac{\tilde U_{mp}(\tilde T)}{\tilde T}
	\right) = k \tilde S_{mp}(\tilde T),
\label{1-3-70}\\
	F_{mp}(T) &=& - k T \ln Z_{mp}(T) = - \frac{4 \pi \hbar c \psi_1}{\tilde{g}^2}
	\tilde T \ln \tilde Z_{mp} (\tilde T) = \frac{4 \pi \hbar c \psi_1}{\tilde{g}^2}
	\tilde F_{mp}(\tilde T).
\label{1-3-80}
\end{eqnarray}
The results of calculations indicate that as $\tilde T\to 0$ there is a regular behavior of the entropy  $\tilde S_{mp}(\tilde T) \xrightarrow{\tilde T \rightarrow 0} \tilde S_1=\text{const.}$ This constant can be set to zero by the corresponding redefinition  of the partition function from \eqref{1-3-30}: $Z_{mp}\to e^{\tilde S_1}Z_{mp}$. The resulting free energy and entropy are shown in Fig.~\ref{entrp_in_enrg_Helmgolz} for the choice $\tilde S_1\approx 3.6$.

In this connection it is interesting to note that, unlike the classical system for  which the entropy demonstrates a singular behavior as  the temperature goes to zero, 
for the nonperturbatively quantized system considered here the entropy is regular as $T\to 0$.

Notice also that since the Helmholtz free energy does not depend on the volume, the pressure $p$ for the system under consideration is equal to zero:
\begin{equation}
	p = - \left(
		\frac{\partial F (T)}{\partial V}
	\right)_T = 0.
\label{1-3-90}
\end{equation}
Using this fact, we can verify our numerical calculations by means of the second law of thermodynamics
\begin{equation}
	T \frac{d S}{d T} = \frac{d U}{d T}.
\label{1-3-100}
\end{equation}
To calculate the left- and right-hand sides of this equation, we have employed numerically computed values of $S(T)$ and $U(T)$.  The results of calculations indicate a perfect agreement of the left- and right-hand sides of \eqref{1-3-100}.
This also confirms that the pressure for the system consisting of one monopole is equal to zero, according to Eq.~\eqref{1-3-90}.

\section{Partition function of a nonrelativistic dilute gas of noninteracting monopoles
}
\label{pfgm}

In this section we consider a nonrelativistic dilute gas of noninteracting monopoles enclosed in the volume $V$. The total energy of such physical system consists of the energy of $n$ monopoles [each of which has the energy defined by Eq.~\eqref{1-1-200}],
of the kinetic energy of the monopoles (which is calculated here in the approximation of point-like particles), and of the energy of quantum condensate with the energy density $\epsilon_{\infty}$:
\begin{equation}
	E_{gas} \Bigl( \vec p_i, (f_2)_i \Bigl) = \sum \limits_{i=1}^n
	\left[
		E_{mp}\Bigl( (f_2)_i \Bigl) +
		\frac{p^2_i}{2 \bigl( m_{mp} \bigl)_i (f_2)_i}
	\right] +
	\epsilon_{\infty} V = E_1\Bigl( \vec p_i, (f_2)_i \Bigl) + E_2.
\label{3-10}
\end{equation}
Here $\bigl( m_{mp} \bigl)_i$ is the mass of $i$-th monopole, $E_1\Bigl( \vec p_i, (f_2)_i \Bigl)$ is the energy of all  $n$ monopoles, and $E_2=\epsilon_{\infty} V$ is the energy of the condensate. To simplify the calculation of its kinetic energy, we assume that the monopole is a point-like particle.
The monopole mass is related to its energy by the expression $m_{mp} = E_{mp}/c^2$. Then the partition function is given by the following integral:
\begin{equation}
	Z_{gas} = \frac{1}{n!} \frac{1}{\left( 2 \pi \hbar \right)^{3 n}}
	\int e^{-\frac{E_{gas} \left( \vec p_i, (f_2)_i \right)}{k T}}
	\prod\limits_{i=1}^n
	dp^i_x dp^i_y dp^i_z dx^i dy^i dz^i d(\tilde f_2)_i =
	\frac{e^{-\frac{\epsilon_{\infty} V}{k T}}}{n!} \left[
		\frac{V}{\left( 2 \pi \hbar \right)^3}\int e^{-\frac{E_{1}(p, f_2)}{k T}} d^3p \; d \tilde f_2
	\right]^n .
\label{3-20}
\end{equation}
In Sec.~\ref{2c} we have calculated the partition function for one quantum monopole assuming that the energy of the quantum condensate, into which the quantum monopole is embedded, does not fluctuate. Continuing working within this assumption, here we calculate the  partition function for $n$ quantum monopoles embedded into the condensate filling the volume $V$.
This means that we \emph{calculate the partition function on the background of the quantum condensate. }

\begin{figure}[h]
	\begin{minipage}[ht]{.45\linewidth}
		\begin{center}
							\includegraphics[width=1\linewidth]{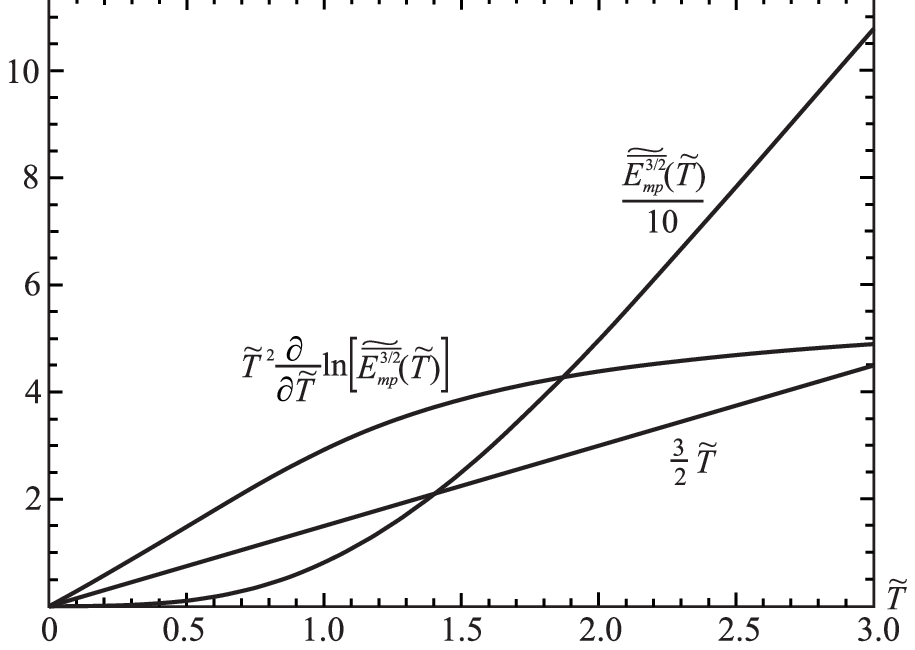}
			\end{center}
		\caption{The dependencies of
		$\tilde T^2
			\frac{
				\partial }{\partial \tilde T}
					\ln \left[ \; \widetilde{\overline{E^{3/2}_{mp}}}
					\Bigl( \tilde T \Bigl)
				\right]$, $\frac{3}{2} \tilde T$, and
		$\widetilde{\overline{E^{3/2}_{mp}}} \Bigl( \tilde T \Bigl)$ on the  temperature	for the monopole gas.
			}
	\label{mpGasInnerEnergy}
	\end{minipage}\hfill
\begin{minipage}[ht]{.45\linewidth}
		\begin{center}
					\includegraphics[width=1\linewidth]{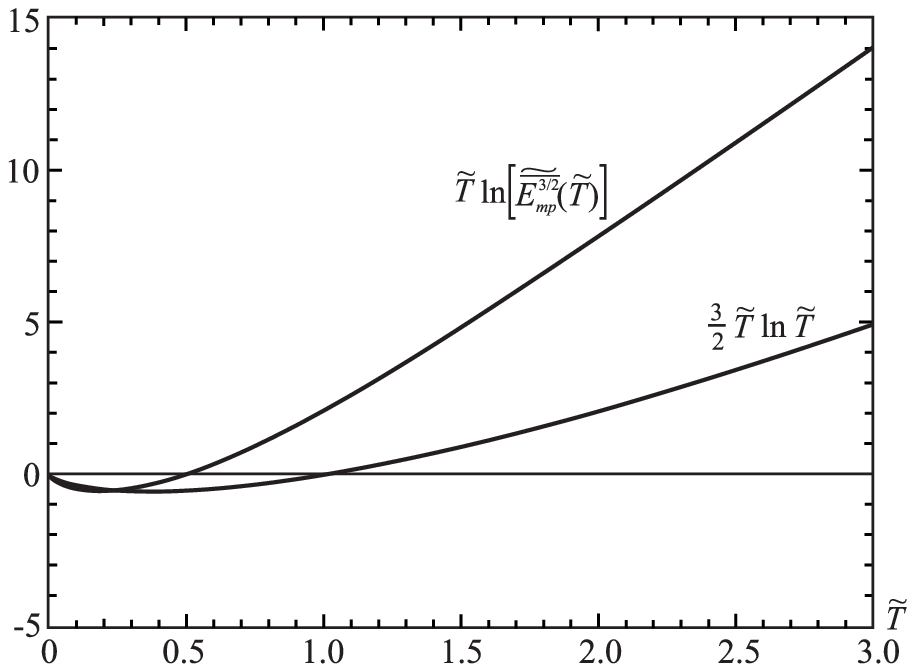}
		\end{center}
		\caption{The dependencies of
		$\tilde T \ln \left[ \; \widetilde{\overline{E^{3/2}_{mp}}}
					\Bigl( \tilde T \Bigl)
				\right]$ and $(3/2) \tilde T \ln{\tilde T}$
on the  temperature  for the monopole gas.
		}
		\label{TlnE32}
\end{minipage}
\end{figure}

Integration over the momentum  $p$ in \eqref{3-20} is performed in the standard way and yields
\begin{equation}
\begin{split}
	Z_{gas} = & \frac{e^{-\frac{\epsilon_{\infty} V}{k T}}}{n!} \left[
		\frac{V}{\hbar^3} \left(
			\frac{k T}{2 \pi c^2}
	\right)^{3/2} \int
		E^{3/2}_{mp}(f_2) e^{-\frac{E_{mp}(f_2)}{k T}} d \tilde f_2
	\right]^n =
	\frac{e^{-\frac{\epsilon_{\infty} V}{k T}}}{n!} \left[
	\frac{V}{\hbar^3} \left(
				\frac{k T}{2 \pi c^2}
		\right)^{3/2}
	\overline{E^{3/2}_{mp}} \Bigl( T \Bigl)
	\right]^n
\\
	&
	=\frac{e^{- \gamma}}{n!} \left[
		\frac{\gamma \left( k T \right)^{5/2}}
		{\left( 2 \pi \right)^{3/2} \hbar^3 \epsilon_\infty c^3}
		\overline{E^{3/2}_{mp}} \Bigl( T \Bigl)
	\right]^n.
\label{3-30}
\end{split}
\end{equation}
Here $\gamma = \frac{\epsilon_{\infty} V}{k T}$ and
\begin{equation}
	\overline{E^{3/2}_{mp}}\Bigl( T \Bigl) = \int E^{3/2}_{mp}(f_2)
		e^{-\frac{E_{mp}(f_2)}{k T}} d \tilde f_2 =
	\left(
		\frac{4 \pi \hbar c \psi_1}{\tilde{g}^2}
	\right)^{3/2}
		\int {\tilde E}^{3/2}_{mp}
		e^{- \frac{\tilde E_{mp}}{\tilde T}} d \tilde f_2 =
	\left(
		\frac{4 \pi \hbar c \psi_1}{\tilde{g}^2}
	\right)^{3/2}
	\widetilde{\overline{E^{3/2}_{mp}}} \Bigl( \tilde T \Bigl)
\label{3-40}
\end{equation}
is the non-normalized average statistical value of the monopole energy $E_{mp}$ which depends on the temperature (here the term ``non-normalized'' means that in defining the average statistical value we do not perform a division by $Z_{gas}$).
Fig.~\ref{mpGasInnerEnergy} shows the dependence of $\widetilde{\overline{E^{3/2}_{mp}}}$ on the dimensionless temperature
$\tilde T$. It is seen that the appearance of the factor $\overline{E^{3/2}_{mp}}\Bigl( T \Bigl)$, which depends on the temperature,
results in a considerable difference between the partition functions of the monopole gas and of a classical monatomic ideal gas for which $Z\sim T^{3n/2}$. Physically this difference is associated with the fact that the monopole is not a point-like particle, and the energy gained from or returned to the thermostat changes not only the kinetic energy of monopoles but their internal energy as well.

Let us now calculate the pressure $p_{gas}$ for the dilute gas of monopoles under consideration
\begin{equation}
\label{3-50}
	p_{gas} = - \left(
		\frac{\partial F_{gas} (T)}{\partial V}
	\right)_T,
\end{equation}
where
\begin{equation}
	F_{gas}(T) = - k T \ln Z_{gas}(T) .
\label{3-60}
\end{equation}
Substituting the expression for $ Z_{gas}$ \eqref{3-30} into \eqref{3-60},
we have from \eqref{3-50}
\begin{equation}
	p_{gas} = - \epsilon_{\infty} + \frac{n k T}{V} .
\label{3-70}
\end{equation}
Comparing this with the expression for a classical monatomic ideal gas, one can
observe the appearance of the  term $- \epsilon_{\infty}$ associated with the energy of the quantum condensate.

In turn, the expression for  the Helmholtz free energy \eqref{3-60} will now be
\begin{equation}
	F_{gas}(T) = \epsilon_{\infty} V -	n k T \left\{
\ln \left[ \;
		\overline{E^{3/2}_{mp}} \Bigl( T \Bigl)
	\right] +
			\ln \frac{V}{V_0} + \frac{3}{2}  \ln \frac{T}{T_0}\right\}.
\label{3-80}
\end{equation}
Here the first and second terms are the contributions coming from the condensate $\phi$ and from a single monopole, respectively;
the last term describes the dilute gas of noninteracting monopoles; $V_0, T_0$
depend on the constants entering $Z_{gas}$. The profile of the dependence of
$\tilde T \ln \left[ \; \widetilde{\overline{E^{3/2}_{mp}}}\Bigl( \tilde T \Bigl)\right]$ on the dimensionless temperature $\tilde T$ is shown in Fig.~\ref{TlnE32}, where the behavior of the fourth term from \eqref{3-80}
is also given for comparison.

Finally, the internal energy is calculated as follows:
\begin{equation}
	U_{gas}(T) = - \frac{\partial \ln Z_{gas}}{\partial \beta} =
		\epsilon_{\infty} V + \frac{4 \pi \hbar c \psi_1}{\tilde{g}^2} n \tilde T^2
		\frac{
			d }{d \tilde T}
				\ln \left[ \; \widetilde{\overline{E^{3/2}_{mp}}}
				\Bigl( \tilde T \Bigl)
			\right] +
		\frac{3}{2} n k T,
\label{3-90}
\end{equation}
where $\beta = 1/(kT)$. The first two terms on the right-hand side of this expression
are the corrections occurring when the internal structure of the monopole is taken into account.
Fig.~\ref{mpGasInnerEnergy} shows the dependence of
$
	\tilde T^2 \frac{\partial }{\partial \tilde T}
	\ln \left[ \; \widetilde{\overline{E^{3/2}_{mp}}}
	\right]
$ on the  temperature $\tilde T$,
and also the third term from the right-hand side of \eqref{3-90} for the sake of comparison.

\section{$\psi_1$ and $\Lambda_{\text{QCD}}$}
\label{connQCDpsi}

In our calculations, there is one undetermined quantity $\psi_1$ having dimensions of
$\text{m}^{-1}$. One can assume that it might be relevant to some physical quantity known from QCD.
Such a quantity is
$\Lambda_{\text{QCD}} = 200 \text{MeV}$, which in units of $\text{m}^{-1}$ is $\Lambda_{\text{QCD}} \approx 10^{15} \text{m}^{-1}$.
This enables us to assume that
\begin{equation}
	4 \pi \hbar c \psi_1 \approx \Lambda_{\text{QCD}}.
\label{4-10}
\end{equation}
In this case the quantum corrections $\left( \mu^2 \right)^{ab \mu \nu}$ and $M^2$ in Eqs.~\eqref{1-1-10} and \eqref{1-1-20}
can be written as
$
	\left( \mu^2 \right)^{ab \mu \nu} =
	\left(\frac{\Lambda_{\text{QCD}}}{\hbar c}\right)^2 \left( \tilde \mu^2 \right)^{ab \mu \nu},
	M = \frac{\Lambda_{\text{QCD}}}{g \hbar c} \tilde M
$, where  $\tilde \mu$ and $\tilde M$ are dimensionless. The relation
 \eqref{4-10} enables us also to assert that $\Lambda_{\text{QCD}}$ describes the dispersion
 $
\left\langle
	\hat A^{m \mu}(y) \hat A^{n \nu}(x)
\right\rangle
$
of quantum fluctuations of the coset fields $\hat A^{m \mu}$.

Then, using $\Lambda_{\text{QCD}}$, the thermodynamic formulae  \eqref{3-40}, \eqref{3-80}, and \eqref{3-90} can be expressed in the following manner:
\begin{eqnarray}
	F_{gas}(T) &=& \epsilon_{\infty} V -
	n \frac{\Lambda_{\text{QCD}}}{\tilde{g}^2} \ln \left[ \;
		\overline{E^{3/2}_{mp}} \Bigl( T \Bigl)
	\right] -
			n k T \ln \frac{V}{V_0} - \frac{3}{2} n k T \ln \frac{T}{T_0},
\label{4-20}\\
	U_{gas}(T) &=&
	\epsilon_{\infty} V + \frac{\Lambda_{\text{QCD}}}{\tilde{g}^2} n \tilde T^2
	\frac{
		\partial }{\partial \tilde T}
			\ln \left[ \; \widetilde{\overline{E^{3/2}_{mp}}}
			\Bigl( \tilde T \Bigl)
		\right] +
	\frac{3}{2} n k T,
\label{4-30}\\
	\overline{E^{3/2}_{mp}}\Bigl( T \Bigl) &=&
	\left(
		\frac{\Lambda_{\text{QCD}}}{\tilde{g}^2}
	\right)^{3/2}
	\widetilde{\overline{E^{3/2}_{mp}}} \Bigl( \tilde T \Bigl).
\label{4-40}
\end{eqnarray}

\section{Conclusion}
\label{concl}

In this work, we investigate the quantization of physical systems in which no small parameter exists that could be used to apply the standard perturbative methods.
In particular, we propose a definition of nonperturbative quantum states in terms of Green functions by using the nonperturbative quantization procedure proposed and developed originally by Heisenberg. We first examine the operator field equations and mention that, due to their mathematical complexity, it is necessary to consider the equivalent system of differential equations that determine the infinite set of Green functions. \emph{Since the general solution of such a system leads to the entire set of Green functions, which must contain all the quantum information about the system, we conclude that we can identify nonperturbative states with Green functions.} We also analyze the case of gauge fields. The corresponding set of infinite equations for the Green functions cannot be solved in general. However, we use earlier results according to which the infinite set of equations can be reduced to a set of two equations whose solutions describe flux tubes corresponding to realistic physical configurations of gauge fields. As a particular example, we study the thermodynamics of one quantum monopole and of a gas of noninteracting quantum monopoles embedded into the condensate.

Working within the framework of the two-equation approximation method for the nonperturbative quantization \`{a} la  Heisenberg, we have found the energy spectrum of one quantum monopole, using which the partition function and the corresponding thermodynamic quantities have been computed. Then we have calculated the partition function and thermodynamic quantities for the system of noninteracting quantum monopoles embedded into the condensate. It was shown that all the thermodynamic quantities have quantum corrections related to the internal structure of the monopole. Physically this means that in the presence of fluctuations the energy received from a thermostat changes not only the kinetic energy of the monopoles but their internal energy as well. This is responsible for the changes in the pressure, internal energy, etc. of the gas of monopoles.

It is important to note that the calculation of statistical and thermodynamic quantities for nonlinear theories quantized by using the nonperturbative methods  \`{a} la  Heisenberg leads to finite results, just as it is in quantum electrodynamics. The principle difference is that in the case considered in the present paper the calculations are carried out for nonperturbatively quantized fields which are not associated with particles (quanta) but are more like a turbulent fluid at each point of which there are fluctuating velocities, pressures, etc.

In summary,  we have
\begin{itemize}
	\item formulated the notion of a nonperturbative quantum state;
	\item approximately defined nonperturbative quantum states for a quantum monopole and for a flux tube
determined by the 2-point Green functions $G^{mn \mu \nu}$,  $G^{ab \mu \nu}$ and the 4-point Green function	$G^{mnpq}_{\phantom{mnpq}\mu \nu \rho \sigma}(x, y, z, u)$;
	\item numerically calculated the energy spectra for a single quantum monopole and for a flux tube;
	\item numerically calculated the partition function and thermodynamic quantities for the quantum monopole and for the dilute gas of noninteracting quantum monopoles.
\end{itemize}
The physical interpretation of the obtained results is that, using the nonperturbative quantization, we have shown that the nonperturbative vacuum in pure gluodynamics (without quarks) consists of a quantum condensate of the coset fields filled with quantum monopoles. This result confirms the hypothesis proposed earlier~\cite{Nambu:1974zg,Hooft,Mandelstam:1974pi}
and confirmed by lattice calculations that the QCD vacuum is filled by monopoles~\cite{Ripka:2003vv}.

\section*{Acknowledgements}

V.D. and V.F. gratefully acknowledge support provided by grant in fundamental research in natural sciences by the Ministry of Education and Science of Republic of Kazakhstan.


\begin{thebibliography}{99}


\bibitem{strocchi13} F. Strocchi, {\it An introduction to nonperturbative foundations of quantum field theory} (Oxford University Press, UK, 2013)

\bibitem{filu15} I. Feranchuk, A. Ivanov, V. Le, and A. Ulyanenkov, {\it Nonperturbative description of quantum systems} (Springer Verlag, Heidelberg, 2015).

\bibitem{dunn16} G. V. Dunne and M. \"Unsal, {\it New nonperturbative methods of quantum field theory: From large-$N$ orbifold equivalenceto bions and resurgence}, Annu. Rev. Nucl. Part. Sci. {\bf 66}, 245 (2016).



\bibitem{hei66} W. Heisenberg, {\it  Introduction to the unified field theory of elementary particles} (Max-Planck-Institut f\"ur Physik und Astrophysik, Interscience Publisher, London, 1966)


\bibitem{Nambu:1974zg}
  Y.~Nambu,
  Phys.\ Rev.\ D {\bf 10}, 4262 (1974).

\bibitem{Hooft}
G. 't Hooft, in: High Energy Physics, edited by A. Zichichi (Editorice Compositori, Bologna, 1975).

\bibitem{Mandelstam:1974pi}
  S.~Mandelstam,
  Phys.\ Rept.\  {\bf 23}, 245 (1976).

\bibitem{Ripka:2003vv}
  G.~Ripka,
  Lect.\ Notes Phys.\  {\bf 639}, pp.1 (2004);
  [hep-ph/0310102].


\bibitem{Shnir:2005xx}
    Y.~M.~Shnir,
  \emph{Magnetic monopoles}
  (Springer, Berlin, Heidelberg, New York,  2005).

\bibitem{DiGiacomo:1999yas}
  A.~Di Giacomo, B.~Lucini, L.~Montesi and G.~Paffuti,
  Phys.\ Rev.\ D {\bf 61} (2000) 034503
    [hep-lat/9906024].

\bibitem{DiGiacomo:1999fb}
  A.~Di Giacomo, B.~Lucini, L.~Montesi and G.~Paffuti,
  Phys.\ Rev.\ D {\bf 61} (2000) 034504
    [hep-lat/9906025].

\bibitem{Chernodub:1997dr}
  M.~N.~Chernodub, F.~V.~Gubarev, M.~I.~Polikarpov and A.~I.~Veselov,
  Prog.\ Theor.\ Phys.\ Suppl.\  {\bf 131}, 309 (1998)
    [hep-lat/9802036].

\bibitem{Chernodub:1997ay}
  M.~N.~Chernodub and M.~I.~Polikarpov,
  In *Cambridge 1997, Confinement, duality, and nonperturbative aspects of QCD* 387-414
  [hep-th/9710205].

\bibitem{Polyakov:1976fu}
  A.~M.~Polyakov,
  Nucl.\ Phys.\ B {\bf 120}, 429 (1977).

\bibitem{Nesti:1996rm}
  F.~Nesti,
  ``Three-dimensional large N monopole gas,''
  hep-th/9610127.

\bibitem{Martemyanov:1997ks}
  B.~V.~Martemyanov and S.~V.~Molodtsov,
  JETP Lett.\  {\bf 65}, 142 (1997)
  [Pisma Zh.\ Eksp.\ Teor.\ Fiz.\  {\bf 65}, 133 (1997)].

\bibitem{Agasian:1997wv}
  N.~O.~Agasian and K.~Zarembo,
  Phys.\ Rev.\ D {\bf 57}, 2475 (1998)
    [hep-th/9708030].

\bibitem{Chernodub:2000mi}
  M.~N.~Chernodub,
  Phys.\ Lett.\ B {\bf 515}, 400 (2001)
    [hep-th/0011124].

\bibitem{Chernodub:2004qp}
  M.~N.~Chernodub, K.~Ishiguro and T.~Suzuki,
  Prog.\ Theor.\ Phys.\  {\bf 112}, 1033 (2004)
    [hep-lat/0407040].

\bibitem{Das:2009fb}
  S.~R.~Das and G.~Murthy,
  Phys.\ Rev.\ Lett.\  {\bf 104}, 181601 (2010)
    [arXiv:0909.3064 [hep-th]].

  \bibitem{Davis:2001mg}
  A.~C.~Davis, A.~Hart, T.~W.~B.~Kibble and A.~Rajantie,
  Phys.\ Rev.\ D {\bf 65}, 125008 (2002)
    [hep-lat/0110154].

\bibitem{Dzhunushaliev:2017rin}
  V.~Dzhunushaliev,
  ``Quantum monopole via Heisenberg quantization,''
  arXiv:1711.01737 [hep-ph].

\bibitem{Dzhunushaliev:2016svj}
  V.~Dzhunushaliev,
  EPJ Web Conf.\  {\bf 138}, 02003 (2017)
  [arXiv:1608.05662 [hep-ph]].

\bibitem{Dzhunushaliev:2015mva}
V.~Dzhunushaliev, V.~Folomeev, B.~Kleihaus and J.~Kunz,
Eur.\ Phys.\ J.\ C {\bf 75}, no. 4, 157 (2015)
doi:10.1140/epjc/s10052-015-3398-5
[arXiv:1501.00886 [gr-qc]].

\bibitem{Dzhunushaliev:2013nea}
V.~Dzhunushaliev, V.~Folomeev, B.~Kleihaus and J.~Kunz,
Eur.\ Phys.\ J.\ C {\bf 74}, 2743 (2014)
doi:10.1140/epjc/s10052-014-2743-4
[arXiv:1312.0225 [gr-qc]].

\end{thebibliography}
\end{document}